\documentclass[twocolumn]{aa}
\usepackage{natbib}
\bibpunct{(}{)}{;}{a}{}{,}
\usepackage{psfig}
\usepackage{epsfig,afterpage} 
\usepackage{graphicx}
\usepackage{pifont}
\usepackage{amssymb}
\usepackage{longtable}
\usepackage{array}
\usepackage{amsmath}
\usepackage{rotating}
\usepackage{multirow}
\usepackage{lscape}
\usepackage{txfonts}

\begin{document}

\title {
        Abundance anomalies in   hot  horizontal branch stars   of the
        Galactic  globular    cluster    NGC~2808.
\thanks{
        Observations collected at the ESO VLT.}}
 
\author{ 
        G. Pace\inst{1},   A.   Recio--Blanco  \inst {2},   G.  Piotto
        \inst{1} \and Y. Momany \inst{1}}
 
\offprints{G. Pace, \email pace@pd.astro.it}

\institute{ 
            Dipartimento di Astronomia, Universit\`a di Padova, Vicolo
            dell'Osservatorio  2,    I-35122 Padova, Italy    \\  \and
            Dpt. Cassiop\'{e}e, UMR  6202, Observatoire de la C\^{o}te
            d'Azur, B.P. 4229, F-06304 Nice Cedex 4, France \\ }


\abstract{ We  present metallicity measurements of   25 stars in the
blue horizontal  branch  of the Galactic  globular cluster NGC~2808.
Our measurements are based on  moderate--resolution spectra taken with
the multi--object  fiber facility FLAMES--UVES,   mounted on Kueyen at
the Very Large   Telescope. We  confirm  that stars hotter  than a
threshold temperature  have super--solar  abundance, while  the cooler
ones respect   the  nominal   metallicity    of the   cluster,    i.e.
[Fe/H]$\simeq$-1.1.    The threshold  temperature is  estimated  to be
about 12\,000~K,   corresponding   to the  so   called  $u$--jump, and
coincides with the sudden departure  of the cluster horizontal  branch
from the models.  The  metallicity increases with temperature for star
hotter than  the   jump, confirming the  hypothesis   that the process
responsible for this abrupt metallic enhancement is the levitation due
to the strong radiation field in absence of a significative convective
envelope.  A metallicity dependence  of  the abundance enhancement  is
also suggested, with more metal poor clusters having a higher increase
in metal content. The    slope in the  temperature vs.    abundance
diagram is  higher than the errors  involved, and the metal content of
the cluster plays possibly a role  in determining the amplitude of the
jump (more metal poor clusters  show more enhancement after the jump),
although other parameters,  such as clusters' characteristics and even
the atomic species involved, may also someway contribute.}

\authorrunning{Pace et al.\ }

\titlerunning{Abundance anomalies in NGC~2808.}

\maketitle

\section{Introduction}
\label{intro}

Stars in the Horizontal branch (HB)  have already passed the red giant
branch   (RGB)  phase  and  are  now   burning  helium in  their  core
\citep{hs55}.  The wide colour  distribution of the  HB, called the HB
morphology, is the result of large differences in the envelope mass of
stars having the same core mass, at  the same evolutionary stage.  HBs
in metal--rich  clusters are usually populated  mostly in the red part
of  the RR--Lyrae instability strip, whereas  metal--poor ones tend to
have  an   extension  on  the  blue   side,   as  first recognised  by
\citet{1stpar}.  It soon  became also clear that clusters  metallicity
is    not  the     only    parameter governing     the   HB morphology
\citep{sw67,vdb67}. As a matter of fact,  in several globular clusters
(GCs) with intermediate and  metal--rich content, a consistent or even
dominant fraction of the HB stars is located blueward of the RR--Lyrae
instability strip, locus to which we refer as BHB for the remainder of
the paper.

Various issues  that   have arisen  about   the HB  of  GCs   from the
observational point of view are  seriously challenging the HB standard
models:
\begin{list}{$\cdot$}{}
\item 
some clusters have HB blue tails that extend very  far to the blue, up
to $\sim$ 30\,000~K or more \citep[see][for a theoretical explanation]
{cc93};
\item 
the stellar  distribution of the HB is  characterised by different {\bf
gaps},   i.e.  different   underpopulated  areas  \citep[see e.g.][and
reference    therein]{p8etal99}  and  luminosity  {\bf  jumps}, i.e.
discrepancies between the observed  and  the modeled  distribution  of
stars in the colour--magnitude diagram \citep{gr99, YAZ1};
\item 
the rotation--velocity  distribution also shows peculiarities: some HB
stars do not rotate as slow as expected for low--mass stars and even a
few fast  rotators  ($v \sin i  \sim 40$  km/sec) have  been  found in
several  clusters  redward  of  the first   luminosity jump  at $\sim$
11\,500~K \citep{peterson83,behr99,rotvel2}
\end{list}

Gradually, some  scenarios  have been  proposed  in order to elucidate
these   facts  which could be     all intertwined \citep[see e.g.][and
reference therein]{behr1}.   Still,  the unresolved  puzzles outnumber
the answered questions.

The HB   luminosity  jump  was  first discovered    in M   13, in   an
intermediate--band photometric study made by  \cite{gr98}.  It was seen
as a sudden departure of the observed location of the  HB stars in the
$(u-y)_0$ vs.    $V$   diagram from  the  model  track.     Soon after
\citep{gr99}, this feature was recognised  to be ubiquitous, occurring
at the same  colour, i.e.  at  the same temperature within the errors,
in  all  the surveyed clusters, no  matter  what their metallicity and
morphology is, provided, of course, that they extend both blueward and
redward of  the jump.  At a  temperature higher than $\sim$ 20\,000~K,
the horizontal  branch position    on the colour--magnitude    diagram
becomes again consistent with the model  track.  This photometric jump
corresponds with that in the T$_{\rm eff}$ vs.   log g diagram of M~15
discovered by \cite{loggjump}.   The  mechanism suggested to  be  most
likely responsible  for both features is  the radiative  levitation of
heavy metals, that  increases  the atmospheric abundance of  the star.
As a result the  opacity due to the metallicity  rises with respect to
the hydrogen  opacity.  Hence there  is  an energy  redistribution  in
which the  Balmer  jump is  partially filled   in, but  the bolometric
emission remains unchanged, and it is therefore seen as an increase in
luminosity in the $u$  Str\"{o}mgren and $U$ Johnson magnitudes  (that
is  why \citet{rollyetal00} find  the gap also in  the  $U-B$ vs.  $U$
diagram).  The occurrence of levitation was already argued by
\cite{lev}.   They  predicted that, if the   outer envelope  of hot HB
stars is  stable enough to  allow helium gravitational settling, which
was hypotesised  by \cite{heset} in  order to  explain observed helium
underabundances,   then   there should be  no    such a mechanism like
convection able  to    prevent the metal  enhancement    caused by the
radiation.   \cite{caloi99} used   radiative levitation  of metals  to
explain the presence  of a gap at   8\,500~K seen in several  clusters
(but   not in NGC~2808).   \cite{hlh00} made the  only  attempt to our
knowledge to model the atmosphere of the HB  stars taking into account
radiative  levitation and  gravitational  settling.  They  managed  to
reproduce  the main observational features,   such as the  photometric
jump,  since  the  luminosity in  each  bandpass depends  on the total
opacity and  not on  the detailed  abundance pattern.   However  these
authors warn that  the element abundances  can only be reproduced with
more sophisticated models.

The advent of the new generation telescopes made  possible to check in
a more direct way whether radiative levitation  is at work and to make
quantitative analysis. Chemical abundances on blue  HB stars have been
more extensively revisited by \cite{behr00a,behr00b}, further extended
in \cite{behr1,  behr2, moehler00, damian}.  New detailed measurements
of  several elements   confirmed   that  blue  HB  stars cooler   than
11\,000-12\,000~K in general  show  no  deviation  from the   globular
cluster abundances derived from red  giants.   On the contrary,  stars
hotter   than 11\,000-12\,000~K,   depart  from  the general  globular
cluster abundances.   In   particular,  iron, titanium   and  chromium
enrichments  to solar  or even   super--solar  values are  detected.  In
addition,   some  metals, such as  phosphorus   and manganese, display
significantly larger enhancements than  iron.  On the other  hand, the
abundances of magnesium, silicon and calcium, both below and above the
critical temperature of 11\,000~K,  are consistent with very little or
no enhancement, even though a  large, likely  real scatter of  silicon
abundance is  observed.  Finally, underabundances  of helium for stars
hotter than 11\,000~K are measured.

Stellar   rotation could  be  tightly  connected  to   the HB chemical
anomalies.  In particular, an abrupt change in the rotational velocity
distribution along the HB at the level of the jump at $\sim$ 11\,500~K
\citep{rotvel1,rotvel2,behr1,behr2}  can possibly be associated to the
onset of radiative levitation.  Angular momentum transfer prevented by
a  gradient in molecular weight \citep[in   a picture a la][]{sp00} or
removal   of  angular     momentum    due  to   enhanced   mass   loss
\citep{rotvel1,vc02} could be responsible  for that discontinuity.  In
any case,  the average rotation rate  is surprisingly high, considered
that braking mechanisms should have spun down the progenitors of these
HB stars, ever  since their  entrance on the   main sequence, and  the
overall observed rotation distribution is far from being understood.

In   summary, the observational   results  of  the  recent years   are
unveiling the  details of a complicated  puzzle of physical parameters
\citep[see e.g.][]{moehler02,ale05} and    open   a number  of     new
questions.   Is there a dependence of  the  chemical enhancements with
the  original metallicity of the   stellar population?  How is stellar
rotation influencing  the chemical  abundance anomalies  or vice versa?
The answer to these  questions will not only  help to our fine--tuning
of stellar  evolution.    Hot  HB stars  play  an   important role  in
population synthesis of  extragalactic non resolved  systems; as these
old   hot stars  can    be easily  confused   with young  hot  stellar
population.  For the same reason, they can also affect the modeling of
the star formation history in dwarf galaxies of  the Local Group which
tries to  reproduce their colour--magnitude diagrams \citep{tolstoy98}.
New  theoretical models with  diffusion and radiative levitation on HB
stars, supported  by    appropriate measurements,  are  essential   to
reconstruct the general picture.

This work is intended to  study the radiative levitation  in the HB of
NGC~2808.  With that purpose, a sample of 32 stars belonging to the HB
of this cluster,  in the range of  effective temperatures from 11\,000
to 16\,000~K, has been observed.   NGC~2808, with its relatively  high
metallicity, \citep[{[}Fe/H{]}=-1.14  according to][]{met2808}, is one
of the most peculiar examples of second--parameter GCs.  It has a very
extended blue tail,  a marked  bimodal distribution \citep{strip2808},
being the loci both redward and blueward of the instability strip very
well     populated   \citep[see][for    a     recent   high--precision
photometry]{rollyetal00}.  The  RR--Lyrae variables are  very  sparse:
\cite{ch89}  found only two,  while 16 of them  could only recently be
found by \cite{2808rrl}   by using  new image--subtraction  techniques
that allowed the  analysis of  central  crowded field.  There  are two
additional  significant gaps in  the  BHB star distribution at 18\,000
and  25\,000~K \citep{sosin97}.  Moreover,  enhancement and  spread of
helium among cluster stars  has been recently  suggested  as a way  to
explain its HB  morphology \citep{dc05}.   Finally, NGC~2808 is  about
20\%    younger      than metal    poor   clusters    of   our  Galaxy
\citep{ros99,franciui}.

Section  2  of this  paper presents   the observations  and   the data
reduction.   In Section 3, the atomic  data  and the derivation of the
atmospheric parameters are addressed.   The  results are presented  in
Section 4   and discussed  in   Section 5 through   a comparison  with
previously analysed clusters.    The related  topic of   the  rotation
velocity will be treated  in a separate  work \citep{rotvel3}, with  a
much wider dataset than the one used  herin, to compensate with a rich
statistic  the  projection of  the rotational  velocity, and with many
more stars cooler  than the luminosity jump,  where most fast rotators
have been found in other clusters.

\section{Observations and    data reduction}    
\label{data}  

We have  observed, in the ESO  observing run 072.D-0742(A), a total of
32 stars,  selected  from the  ground  based wide field photometry  by
\cite{yazzz,YAZphot}.  Due to their  faintness (their visual magnitude
is between 16 and 17.5  mag) and to the  fact that they were collected
within  a more general  programme aimed at  the  measurement of radial
velocity of many   hundreds  of stars  \citep{p8etal04}, some  of  the
spectra  have a limited   S/N ratio, ranging from   few units to a few
tens.   Therefore,  not all  the observed stars  were  suitable  for a
reliable measurement of their abundance.  Our final sample contains 24
stars, 18 of which are hotter than 12\,000~K.   The homogeneity of the
star distribution  in  a   temperature  range  that extends  to   over
16\,000~K and the considerable number of the  targets, make our sample
unique  among those so far observed  in a single  cluster for the same
purpose.

Targets  lie   in the  low--crowded  outskirts   of the  GC,  to avoid
contamination from other stars.  The  spectra were obtained in service
mode in December 2003 and February 2004.  The data have been collected
with the  FLAMES fiber facility connected  to the UVES spectrograph at
the focus  of the Unit  2, Kueyen, of the Very  Large  Telescope.  The
resolution  was of  R=47\,000  and the  exposure  times ranged between
2\,400  s  and 3\,000  s.   The  spectra have   been reduced  with the
FLAMES--UVES pipeline  \citep{flamesuves}, and then  analysed both with
MIDAS and IDL routines.

\section{Abundance measurements}

The abundance measurements   were performed using the program  WIDTH3,
developed by R. G.  Gratton and adapted by  D. Fabbian to temperatures
up to  $T_{\rm  eff}  \simeq$ 20\,000~K \citep{damiant,damian}.    The
procedure  establishes  the    abundances   of  chemical species    by
reproducing the  observed equivalent widths.   Once a  starting set of
values for the effective temperature, surface gravity, and model metal
abundance is derived,  an  appropriate  spectrum   for each star    is
obtained  from the  grid of model   atmospheres by \cite{kurucz94}  by
interpolating linearly in the  temperature and logarithmically  to get
the Rosseland opacity,  electronic and gaseous pressure,  and density.
Continuum opacity  was obtained   taking  into account  all  important
continuum   opacity   sources   for    stars    as  hot   as   $T_{\rm
eff}=20\,000$~K. Collisional  damping constants were computed with the
Uns$\ddot{\rm  o}$ld  formula.  The   equation of   transfer was  then
integrated through the atmosphere  at different wavelengths along  the
line profile and theoretical  equivalent widths of the lines  computed
and   compared    with the   observed    ones  \citep[see][for  further
details]{damian}.

\subsection{Oscillator strengths, line selection, and solar abundances}

Only lines clean  from  blends were considered   in the analysis.   In
order to  properly identify the spectral  lines,  we have  used a line
list   suitable for our   stars.   The  list  assembled  and  used  by
\cite{damian}  was    completed, for  wavelength  values   larger than
5\,000~\AA,   thanks to  the     DREAM line data  base  \citep{DREAM},
accessible via  VALD \citep[][and references therein]{VALD}.   For the
determination of the  oscillator  strengths  ({\it gf}s),   laboratory
values were considered  whenever possible.  For \ion{Fe}{i} lines they
were taken from papers  of the  Oxford  group \citep{oxg1,oxg2}.   The
{\it gf}s for the \ion{Fe}{ii} lines were taken from \cite{morefe} and
\cite{b91}. The solar abundances were taken from \cite{solarab}.

\subsection{Equivalent Widths} 

The use of  a  FORTRAN  program  (courtesy  of Patrick Francois)    to
automatically  detect lines and   measure their equivalent widths  was
attempted,  but  due once  again  to the  rather low S/N  ratio of the
spectra, a more interactive procedure turned out to be more effective.
Namely, we  have manually detected    the presence of lines  and  then
fitted those  lines that were  identified (see  next  Section) with  a
Gaussian.  We have then  finally  measured the equivalent width.   The
two latter steps were performed with IDL.

Due to the quality  of the spectra, the  $EW$\ uncertainties  were the
most important source of errors in  the abundance determinations. This
error was evaluated by comparing the  equivalent width measurements of
the \ion{Fe}{ii}  line at 4\,923.93~\AA\ in  the five following stars:
37456, 41077, 41388, 46386 and 8072, namely, the stars before the jump
for  which this  line was detected.    These stars  have almost  equal
atmospheric parameters, with differences just slightly larger than the
internal errors.    An analysis  of  the  line  measured  in  the five
spectra, indicates that the standard deviation of the equivalent width
measurements is $12.6$~m\AA.  We    expect this value    to marginally
overestimate  the equivalent width   errors, since  the stars Are  not
identical.  The value  calculated through the \cite{cayrel88} formula:
$$\Delta(EW)  = 1.6 \sqrt{(w   dx)}/(S/N)$$ with $w$\  being  the full
width  at  half maximum  (FWHM)  typical of  the  lines,  in this case
$\sim$0.12~\AA, and $dx$\ being the pixel size, about 0.03~\AA, we get
$\Delta(EW)\simeq 10$~m\AA (assuming  a $S/N\simeq$10).  As  expected,
this value is lower  than the standard deviation   of the line  in the
five  stars  considered, but only   by  a small  amount.   This result
implies that identification of   the correct continuum level (that  is
neglected  in the Cayrel formula)  is not  an  issue here, as expected
since the spectra of blue HB stars have very few absorption lines.

Consequently,   we calculated  the abundance  errors,  derived from an
equivalent width  uncertainty  of 10.4~m\AA.  

\begin{figure}[ht]
\centering
\includegraphics[width=8cm]{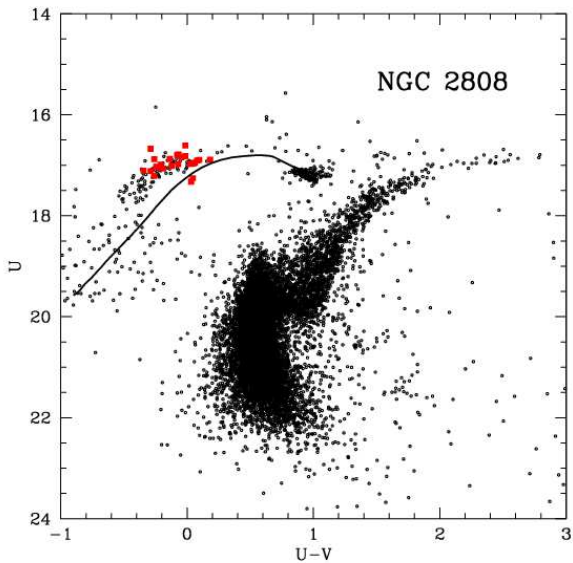}
\caption{
The  {\it U,  (U-V)} diagram of  NGC  2808 from \cite{YAZphot} and the
ZAHB  model   from \cite{ZAHBmodel}. The  target  stars  Are marked as
filled squares.  }
\label{fig1}
\end{figure}

\subsection{Atmospheric parameters} 

\begin{table*}
\begin{center}
\begin{tabular}{c c c c c c c c}
\hline
\hline  
  STAR  & R.A. & dec &  U   &   B   &  V    &  $T_{eff_{U-V}}$&  $T_{eff_{U-B}}$\\
        &      &     &\multicolumn{3}{c}{[mag]} &\multicolumn{2}{c}{[K]} \\
\hline  
   8072 & 9:11:52.3843 & -64:49: 3.953 & 17.25 & 17.39 & 17.21 & 11684 & 12113 \\ 
  10277 & 9:12:32.7088 & -64:46:58.837 & 16.98 & 17.32 & 17.19 & 14725 & 15077 \\ 
  13064 & 9:11:35.8157 & -64:49:42.433 & 17.32 & 17.50 & 17.29 & 11802 & 12587 \\ 
  14923 & 9:11:27.8279 & -64:46:22.480 & 17.21 & 17.58 & 17.47 & 15475 & 15738 \\ 
  31470 & 9:11:42.4870 & -64:54: 1.044 & 17.06 & 17.34 & 17.25 & 14400 & 13985 \\ 
  37289 & 9:11:50.1411 & -64:57:33.778 & 16.89 & 16.95 & 16.80 & 11218 & 11201 \\ 
  37456 & 9:11:58.2345 & -64:57:13.008 & 16.97 & 17.05 & 16.92 & 11645 & 11444 \\ 
  39354 & 9:12:25.9222 & -64:55:12.385 & 16.84 & 17.00 & 16.89 & 12603 & 12253 \\ 
  39744 & 9:11:46.3567 & -64:54:57.860 & 17.11 & 17.46 & 17.40 & 15943 & 15314 \\ 
  40169 & 9:11:44.7537 & -64:54:44.000 & 17.01 & 17.23 & 17.13 & 13514 & 13179 \\ 
  41077 & 9:12:7.7814  & -64:54:43.751 & 16.90 & 16.94 & 16.82 & 11363 & 10947 \\ 
  41388 & 9:12:10.3543 & -64:54:10.584 & 16.95 & 16.99 & 16.88 & 11531 & 11060 \\ 
  45309 & 9:11:50.5199 & -64:52:56.754 & 17.03 & 17.34 & 17.27 & 15186 & 14504 \\ 
  46386 & 9:12:12.7009 & -64:52:38.390 & 16.97 & 17.08 & 16.94 & 11862 & 11683 \\ 
  47485 & 9:11:46.2764 & -64:52:20.187 & 16.88 & 17.19 & 17.14 & 15475 & 14544 \\ 
  47759 & 9:11:46.9302 & -64:52:16.032 & 16.79 & 16.96 & 16.87 & 12971 & 12440 \\ 
  49334 & 9:12:19.4986 & -64:51:51.615 & 16.82 & 17.09 & 16.83 & 12275 & 14060 \\ 
  53685 & 9:12:17.6106 & -64:50:42.834 & 16.61 & 16.75 & 16.62 & 12275 & 12210 \\ 
  53926 & 9:12:5.4170  & -64:50:39.279 & 16.79 & 17.00 & 16.85 & 12831 & 13033 \\ 
  54373 & 9:11:52.5314 & -64:50:32.329 & 16.98 & 17.23 & 17.05 & 12900 & 13585 \\ 
  54411 & 9:12:0.1819  & -64:50:31.816 & 16.67 & 17.01 & 16.96 & 15943 & 15120 \\ 
  55983 & 9:12:15.3646 & -64:50: 4.984 & 17.11 & 17.47 & 17.45 & 16914 & 15647 \\ 
  56118 & 9:12:10.9329 & -64:50: 2.593 & 16.87 & 17.10 & 17.01 & 13699 & 13311 \\ 
  56998 & 9:12:10.2714 & -64:49:47.545 & 17.04 & 17.31 & 17.25 & 14665 & 13911 \\ 
\hline
\end{tabular}
\end{center}
\caption{
         The  photometric  data   adopted  to  estimate the  effective
         temperatures of the stars and the relative outcomes. Both U-V
         and B-V  colours can be used,  obtaining the results shown in
         the respective columns.  
         }
\label{tabphot}
\end{table*}

\begin{table}
\begin{center}
\begin{tabular}{c c c c }
\hline
\hline
STAR     &  $T_{\rm eff}$         &  $\log g$         & $\xi$          \\
         &     [K]            &                   & [km/sec]       \\
\hline
   8072 &    11899 $\pm$  294 &  3.99 $\pm$  0.04 &  1.75 $\pm$ 1 \\ 
  10277 &    14901 $\pm$  289 &  4.25 $\pm$  0.03 &  0.0+1.0-0.0  \\
  13064 &    12194 $\pm$  501 &  4.01 $\pm$  0.05 &  0.0+1.0-0.0  \\ 
  14923 &    15606 $\pm$  265 &  4.31 $\pm$  0.03 &  0.0+1.0-0.0  \\ 
  31470 &    14193 $\pm$  354 &  4.19 $\pm$  0.03 &  0.0+1.0-0.0  \\ 
  37289 &    11209 $\pm$   77 &  3.91 $\pm$  0.03 &  1.87 $\pm$ 1 \\ 
  37456 &    11545 $\pm$  176 &  3.95 $\pm$  0.03 &  1.81 $\pm$ 1 \\ 
  39354 &    12428 $\pm$  244 &  4.04 $\pm$  0.03 &  0.0+1.0-0.0  \\ 
  39744 &    15628 $\pm$  487 &  4.31 $\pm$  0.04 &  0.0+1.0-0.0  \\ 
  40169 &    13346 $\pm$  306 &  4.12 $\pm$  0.03 &  0.0+1.0-0.0  \\ 
  41077 &    11155 $\pm$  284 &  3.91 $\pm$  0.04 &  1.88 $\pm$ 1 \\ 
  41388 &    11295 $\pm$  307 &  3.92 $\pm$  0.04 &  1.85 $\pm$ 1 \\ 
  45309 &    14845 $\pm$  479 &  4.25 $\pm$  0.04 &  0.0+1.0-0.0  \\ 
  46386 &    11773 $\pm$  168 &  3.97 $\pm$  0.03 &  1.77 $\pm$ 1 \\ 
  47485 &    15009 $\pm$  626 &  4.26 $\pm$  0.05 &  0.0+1.0-0.0  \\ 
  47759 &    12705 $\pm$  479 &  4.06 $\pm$  0.04 &  0.0+1.0-0.0  \\ 
  49334 &    13168 $\pm$  991 &  4.11 $\pm$  0.09 &  0.0+1.0-0.0  \\ 
  53685 &    12243 $\pm$  134 &  4.02 $\pm$  0.03 &  0.0+1.0-0.0  \\ 
  53926 &    12932 $\pm$  195 &  4.08 $\pm$  0.03 &  0.0+1.0-0.0  \\ 
  54373 &    13243 $\pm$  453 &  4.11 $\pm$  0.04 &  0.0+1.0-0.0  \\ 
  54411 &    15531 $\pm$  572 &  4.30 $\pm$  0.04 &  0.0+1.0-0.0  \\ 
  55983 &    16280 $\pm$  864 &  4.36 $\pm$  0.05 &  0.0+1.0-0.0  \\ 
  56118 &    13505 $\pm$  371 &  4.13 $\pm$  0.03 &  0.0+1.0-0.0  \\ 
  56998 &    14288 $\pm$  549 &  4.20 $\pm$  0.04 &  0.0+1.0-0.0  \\ 
\end{tabular}
\end{center}
\caption{
  Atmospheric parameters used to carry out the abundance analysis of our
  sample stars.  For the initial guess on  the metallicity the cluster's
  iron abundance,  i.e.  [Fe/H]=-1.14~dex, has  been  used for the stars
  before the jump, and the solar value for the others. These values have
  been iteratively changed  until     the  achievement of  the     final
  convergence within 0.01 dex.}
\label{tab1}
\end{table}

Model atmospheres  appropriate for each star   were extracted from the
grid  of \cite{kurucz94}.  $T_{\rm  eff}$\ values were determined from
the colour of  the  stars in the  Johnson  $U$ vs. $(U-V)$  diagram by
\citet[][see  Fig.  1 therein]{YAZphot} and in  the  a $U$ vs. ($U-B$)
diagram of the same authors.  The models of Zero Age Horizontal Branch
(ZAHB)  by  \cite{ZAHBmodel}  were  used.    Hereinafter, we adopt   a
reddening value of $E(B-V)=0.22$, from the \cite{harris96} compilation
in its revised version of 2003.

\begin{figure}[h]
\centering
\includegraphics[width=8cm]{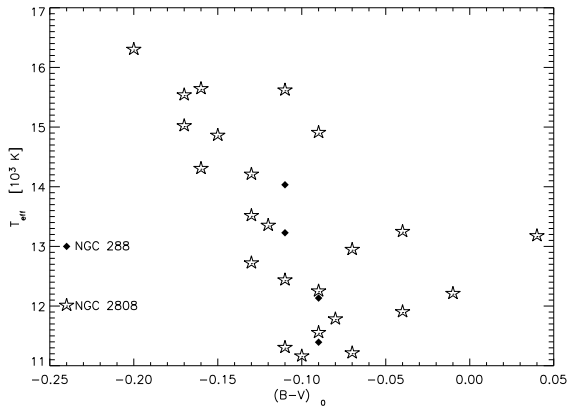}
\caption{
Comparison  between the temperatures derived  for  NGC~2808 and
those of Behr for  NGC~288, chosen because of the similar metallicity.}
\label{fig2}
\end{figure}

To  derive the final effective  temperature of the programme stars, we
calculated  the mean of the  derived temperatures  using the (U-V) and
the (U-B)  colours.  Table~1 presents   the photometric parameters for
all the observed stars in NGC~2808, with the corresponding temperature
values  derived from the  colour--magnitude diagram of \cite{YAZphot},
labelled $T_{eff_{U-V}}$ and $T_{eff_{U-B}}$.   Internal errors in the
adopted  temperatures  were derived  in the   following way:  we  have
derived  four  values  of the  temperature, two   for  each of the two
photometries, respectively adding  and subtracting  the errors on  the
colours. We have then computed the half of  the difference between the
largest and   the  smallest values  obtained.  Clearly,   this kind of
procedure relies on present  ZAHB models without radiative  levitation
unable to reproduce  the observed horizontal  branches between T$_{\rm
eff}  \sim $ 12\,000~K and  20\,000~K.  This introduce external errors
that are  difficult to   evaluate.   In  particular, the   temperature
determination relies on the criterion  adopted to choose the point  on
the model track  that represents our  sample star, which, if enhanced,
is  far from  being on  the track.  In   the present  analysis we have
taken, for each star, the point on  the model track  that has the same
colour as  the star, according to  the fact that temperatures are more
colour sensitive  in this region.   Another approach is to  choose the
closest point of the ZAHB to the star.  On the  other hand, one of the
objectives of this work is also to help  the developing of the models.
Nevertheless, in order  to compare the  effective temperatures adopted
throughout this paper with  those adopted in other  abundance analysis
of blue HB stars, we plotted (see Fig.~\ref{fig2}) the runs of $T_{\rm
eff}$\  against the  dereddened Johnson  $B-V$\  colour  for  both our
program stars in NGC~2808 and the stars in NGC~288 analyzed by
\cite{behr1}.  NGC~288 was chosen because it  has the same metallicity
as    NGC~2808.   The   reddenings   here   adopted    are  those from
\cite{harris03} compilation.

As  in  \cite{damian},  gravities are not    well constrained  by  our
spectra:  in fact  equilibrium of ionization  is  subject to  possible
departures from LTE,  and the Balmer lines are  too broad for reliable
determinations of their profile from our echelle spectra. We have then
determined the gravities of  our target stars from their corresponding
effective temperatures, using the  same  method that \cite{damian}:  a
mean relation between $\log T_{\rm eff}$\ and gravity ($\log g$), from
\cite{behr1}  measurements  of blue HB   stars in  NGC~288,  that were
expected to be very similar to  those in NGC~2808 on  the basis of the
colour--magnitude diagram:

\begin{equation} 
\log g = 2.72 \cdot \log T_{\rm eff} - 7.10 
\label{loggequation}
\end{equation} 

Errors  in  the gravities have  been derived  from  the errors  in the
temperatures using Equation \ref{loggequation}.

Microturbulent velocities $\xi$ might be derived by eliminating trends
of  the derived abundances with  expected line strength for some given
species. However, given the quality of our  spectra, generally too few
lines could be measured, and  with too large scatter, to significantly
constrain microturbulent velocity.  For the cooler stars, we relied on
Behr's analysis of M~13  stars (as no cool HB  stars have been already
analysed  in  NGC~288). From  his  analysis, we derived  the following
relation between $\log T_{\rm eff}$\ and the microturbulence velocity:
\begin{equation} 
\xi = -4.7 \cdot \log T_{\rm eff} + 20.9 ~~~{\rm km/s} 
\end{equation} 

Uncertainties in these values for the microturbulent velocities can be
obtained from  the   scatter of  individual   values around this  mean
relation: this is  $\sim 1$~km/s. For  the warmer stars, which  have a
very  stable atmosphere, still enough  to   let helium sink and  heavy
metals levitate, we adopted 0 microturbulent velocity, in agreement
with the values derived by \cite{behr1}  in  NGC~288. 

\begin{figure}[!ht]
\begin{center}
\resizebox{8cm}{8cm}{
\begin{tabular}{c}
\includegraphics[width=8cm]{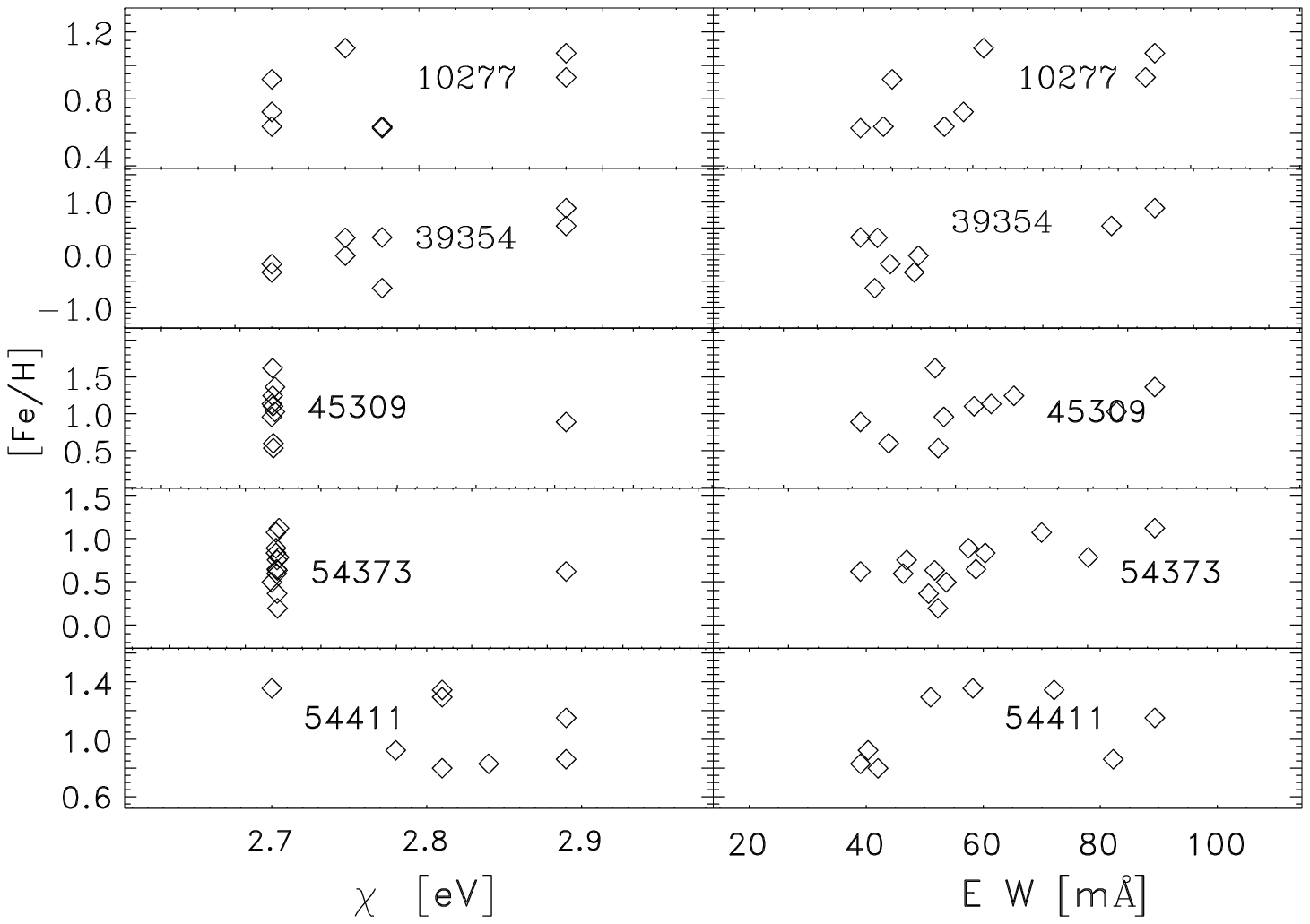}\\
\includegraphics[width=6cm]{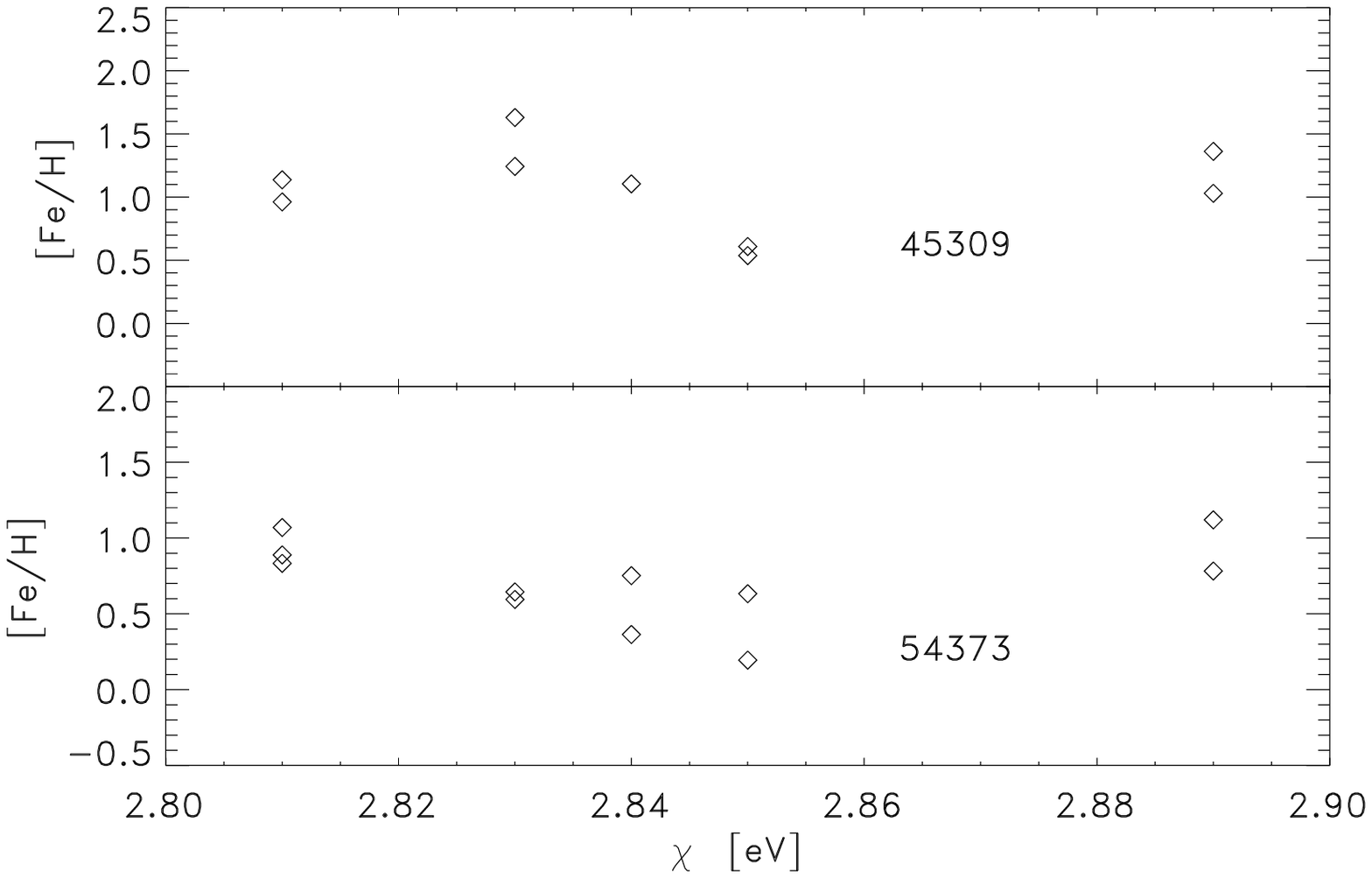}\\
\end{tabular}
}
\end{center}
\caption{
Check of  the accuracy of the photospheric parameters  chosen for the
chemical  analysis. The lines that  on the top--panel graphics appear
crowded are zoomed in the bottom panel.}
\label{fig3}
\end{figure}

A check on the accuracy of  the chosen atmospheric parameters is given
in Figure~\ref{fig3}.  The plot  of excitation potential ($\chi$)  vs.
$[Fe/H]$ and Equivalent Width($EW$) vs. $[Fe/H]$  is shown for all the
stars with eight or more \ion{Fe}{ii} line detections.  The number of
\ion{Fe}{ii} lines  is never bigger than 13,  hence we cannot use this
method,   as  already stated,   to   determine  directly   atmospheric
parameters.  But  the fact than  in none  of  the 5  stars analysed  a
significant slope   appears, suggests that   $\xi$  and $\log  g$  are
properly chosen.  This  is actually a check  on the $\xi$=0 assumption
for the stars  after   the jump,  since  all  of  the five   stars  in
Figure~\ref{fig3} are on its blue side.

Finally, metallicities were obtained varying the metal abundance [A/H]
of the model until it was close enough, namely within 0.01~dex, to the 
derived [Fe/H] value.

Table~2 presents the atmospheric parameters of the analysed stars with
their corresponding  internal  errors. Once  $T_{\rm eff}$,  $\log g$,
$\xi$\ and atmospheric [Fe/H] were  determined, we calculated the mean
abundance and dispersion for   each element. Since the abundances  for
each  element depend upon the  adopted parameters, we recalculated the
abundances while varying each of these parameters in turn, in order to
test the stability of our results.

\section{Results}

Table~\ref{tab3} contains   the   abundances of iron,   titanium,  and
chromium derived  for each   of our  target  stars.  The   values  are
relative to the solar abundances of \cite{solarab}. The scatter of the
measurements    for the corresponding      lines  gives the   quantity
$\sigma_{N}$, which is added quadratically  to the error contributions
in  $T_{\rm  eff}$, $g$, $\xi$,  and  [A/H]  described in the previous
Section, to get the final error for each element  in each target star.
The values for the  dominant ionization stage  (usually singly--ionized
metals) were  used   as  a best   indicator  of  the  actual  chemical
abundance,  since these   stages  are  least  susceptible to   non-LTE
effects.

\begin{figure}[h]
\centering
\includegraphics[width=8cm]{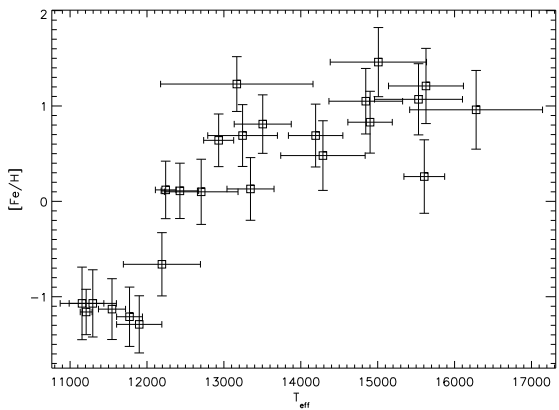}
\caption{
Apparent discontinuity in the [Fe/H] vs.  T$_{\rm eff}$ diagram: stars
hotter  than 12\,000~K have metallicities  more than 1 dex higher than
the  cooler ones. The increasing  metallicity trend of the metallicity
with temperature after the jump is also evident.}
\label{fig4}
\end{figure}

Figure~\ref{fig4} shows  the  iron abundances  as  a  function of  the
effective  temperature  derived for our targets  in   NGC~2808.  It is
clear  that a pattern of  radiative levitation is  present for blue HB
stars,  with  iron   abundances increasing gradually   as  temperature
increases along  the HB, up to a   value of [Fe/H]  $\simeq +1.0$ dex.
One   more remarkable feature, not  so  clearly observed in previously
analysed  clusters,  is the smooth   ascending  metallicity trend with
temperature  after the  jump:   the levitation becomes  more  and more
efficient   at high    temperatures, hence    for  stronger  radiation
fields. Such a trend could only be  discovered because of the richness
of  our sample and the uniformity   of the temperature distribution of
the stars in a very wide range, that  goes from $\sim$ 1\,000~K before
the jump up to over  16\,000~K.  These qualities are unprecedented for
single--cluster samples  observed in this kind  of studies.  The trend
can be either real or result from an  incorrect temperature scale.  On
the one  hand, there are to  date no available  model atmospheres that
take  into account radiative levitation.   The discrepancy between the
predictions of the  model and the photometric  data increases with the
temperature, as clearly noticeable in Figure~\ref{fig1}. Consequently,
abundance  overestimations could  be   stronger  for stars   at larger
temperatures, thus  producing a spurious   trend in the abundance  vs.
$T_{\rm eff}$ diagram.   On the other hand, the  slope that we find in
Figure~\ref{fig4} is too large to be  easily dismissed as an effect of
systematic errors and    it has, on  the contrary,   a straightforward
physical  explanation:    stronger   radiation fields   cause   higher
enhancements. We will not be able to say the  final word on this until
models for hot HB stars of stars are refined.

Although the overall trend, whether real  or artefact, is unmistakably
apparent, the spread of the  data points about  it is not marginal and
there two strong outliers: 14923 at $T_{\rm eff}$=15\,606 and 49334 at
$T_{\rm eff}\sim$=13\,168, which are respectively over 1 dex above and
almost 1 dex below stars  with similar temperatures.  The reason could
reside in the measurement errors, as they are both stars with only one
line   detected.  In addition,   49334 has also  a large  error in the
temperature evaluation.  We also have to remind ourselves that for the
hottest stars the non--LTE behaviour of the stellar atmosphere is more
and more important,  hence our measurements,  based on  LTE modelling,
less precise.  This could lead to spurious  trends for the few hottest
stars  of  the sample,  for  which one would  be  tempted to suggest a
saturation effect.

Looking at Figure~\ref{fig4} we notice, in addition, that the onset of
levitation is quite abrupt for NGC 2808: in less than 500~K there is a
change of more than 1 dex in Fe.

On the other hand, as it can be  seen from Figure~\ref{fig4}, the iron
abundances for the   coolest  stars are   very close  to  the  cluster
metallicity obtained here from analysis of red giant branch stars.  We
obtain an average value  of [Fe/H]=$-1.16\pm 0.08$, that is compatible
with the abundances obtained from analysis of red giants: [Fe/H]=-1.14
on the  \cite{CG} scale (which  should be consistent  with the present
analysis,  as  the    code  of  \cite{damian}   for   our  metallicity
measurements is  the extension of  the software  used in the \cite{CG}
paper);   and the  slightly   lower  value  of  [Fe/H]=-1.23 given  by
\cite{ki03}.

\begin{table*}
\begin{center}
\begin{tabular}{c c c c c c c c c c}
\hline
\hline
Star&[Fe/H]&$\sigma_{lines}$&$N_{lines}$&$\Delta_T$&$\Delta_{logg}$&$\Delta_{[A/H]}$&%
$\Delta_{\xi}$&$\Delta_{EW}$&$\Delta_{tot}$\\
\hline
 8072 & -1.29 & 0.10 &  2 & 0.03   & 0.01    &$<$0.01&  0.19  & 0.23  &  0.30  \\
10277 &  0.83 & 0.20 &  8 & 0.10   & $<$ 0.01&  0.02 &  0.07  & 0.30  &  0.32  \\
13064 & -0.66 & 0.08 &  2 & 0.05   & 0.01    &$<$0.01&  0.07  & 0.32  &  0.33  \\
14923 &  0.26 &  --  &  1 & 0.10   & $<$ 0.01&  0.02 &  0.09  & 0.36  &  0.39  \\
31470 &  0.69 & 0.30 &  5 & 0.10   & $<$ 0.01&$<$0.01&  0.09  & 0.30  &  0.33  \\
37289 & -1.16 &  --  &  1 & $<$0.01& 0.01    &$<$0.01&  0.06  & 0.23  &  0.24  \\
37456 & -1.13 &  --  &  1 & 0.02   & 0.01    &$<$0.01&  0.23  & 0.22  &  0.32  \\
39354 &  0.11 & 0.49 &  8 & 0.03   & 0.01    &$<$0.01&  0.07  & 0.28  &  0.29  \\
39744 &  1.21 & 0.43 &  5 & 0.19   & $<$ 0.01&  0.06 &  0.04  & 0.34  &  0.39  \\
40169 &  0.13 & 0.11 &  2 & 0.07   & 0.01    &  0.02 &  0.11  & 0.30  &  0.33  \\
41077 & -1.07 & 0.70 &  2 & 0.04   & 0.01    &  0.01 &  0.30  & 0.23  &  0.38  \\
41388 & -1.07 & 0.02 &  2 & 0.04   & 0.01    &$<$0.01&  0.28  & 0.21  &  0.35  \\
45309 &  1.05 & 0.33 & 10 & 0.15   & $<$ 0.01&  0.02 &  0.06  & 0.30  &  0.34  \\
46386 & -1.21 & 0.05 &  2 & $<$0.01& $<$ 0.01&$<$0.01&  0.22  & 0.22  &  0.31  \\
47485 &  1.46 & 0.24 &  7 & 0.20   & $<$ 0.01&  0.01 &  0.11  & 0.28  &  0.36  \\
47759 &  0.10 & 0.31 &  4 & 0.09   & $<$ 0.01&  0.02 &  0.11  & 0.31  &  0.34  \\
49334 &  1.23 &  --  &  1 & 0.03   & 0.01    &  0.02 &  0.05  & 0.28  &  0.29  \\
53685 &  0.12 & 0.21 &  3 & $<$0.01& $<$ 0.01&  0.02 &  0.11  & 0.28  &  0.30  \\
53926 &  0.64 & 0.04 &  3 & 0.04   & $<$ 0.01&  0.01 &  0.11  & 0.25  &  0.28  \\
54373 &  0.69 & 0.26 & 13 & 0.09   & $<$ 0.01&$<$0.01&  0.08  & 0.30  &  0.32  \\
54411 &  1.07 & 0.24 &  8 & 0.19   & $<$ 0.01&  0.02 &  0.07  & 0.31  &  0.37  \\
55983 &  0.96 & 0.13 &  2 & 0.26   & $<$ 0.01&  0.02 &  0.08  & 0.31  &  0.41  \\
56118 &  0.81 & 0.30 &  7 & 0.10   & $<$ 0.01&$<$0.01&  0.08  & 0.28  &  0.31  \\
56998 &  0.48 & 0.15 &  4 & 0.16   & $<$ 0.01&  0.02 &  0.07  & 0.32  &  0.37  \\
\hline
         &[Ti/H]&      &     &           &         &       &        &       &       \\
\hline
  39354  & 1.30 & 0.76 &  3  &    0.11   &$<$ 0.01 & 0.01  &   0.08 &  0.30 & 0.32  \\
  40169  & 2.01 & --   &  1  &    0.11   &$<$ 0.01 &$<$0.01&   0.11 &  0.33 & 0.36  \\
  54373  & 1.13 & 0.23 &  5  &    0.17   &$<$ 0.01 & 0.03  &   0.05 &  0.30 & 0.35  \\
  56118  & 1.00 & --   &  1  &    0.15   &$<$ 0.01 & 0.04  &   0.03 &  0.29 & 0.33  \\
\hline
         &[Cr/H]&      &     &           &         &       &        &       &       \\
\hline
  31470  & 0.11 &  --  &  1  &    0.12   &$<$ 0.01 & 0.05  &   0.02 &  0.45 & 0.47  \\
  40169  & 1.19 &  --  &  1  &    0.09   &$<$ 0.01 & 0.01  &   0.10 &  0.33 & 0.36  \\
  41388  &-0.93 &  --  &  1  &    0.04   &    0.01 & 0.01  &   0.03 &  0.48 & 0.48  \\
  49334  & 0.68 &  --  &  1  &    0.32   &$<$ 0.01 & 0.03  &   0.05 &  0.29 & 0.44  \\
\hline
\end{tabular}
\end{center}
\caption{Result of  the  abundance   analysis of iron,   chromium  and
         titanium and relative errors  resulting from the  1\,$\sigma$
         uncertainties in the  assumed parameters.  In the last column
         the resulting 1\,$\sigma$ error is indicated. The quantity in
         the $\sigma_{lines}$  column  is the standard  deviation when
         the number of lines detected is larger than 4. When there are
         only 4 lines or less,  the value reported  in the same column
         is  half  the  difference  between  the largest   and  lowest
         measured abundances.}
\label{tab3}
\end{table*}

\begin{figure}[h]
\centering
\begin{tabular}{c}
\includegraphics[width=8cm]{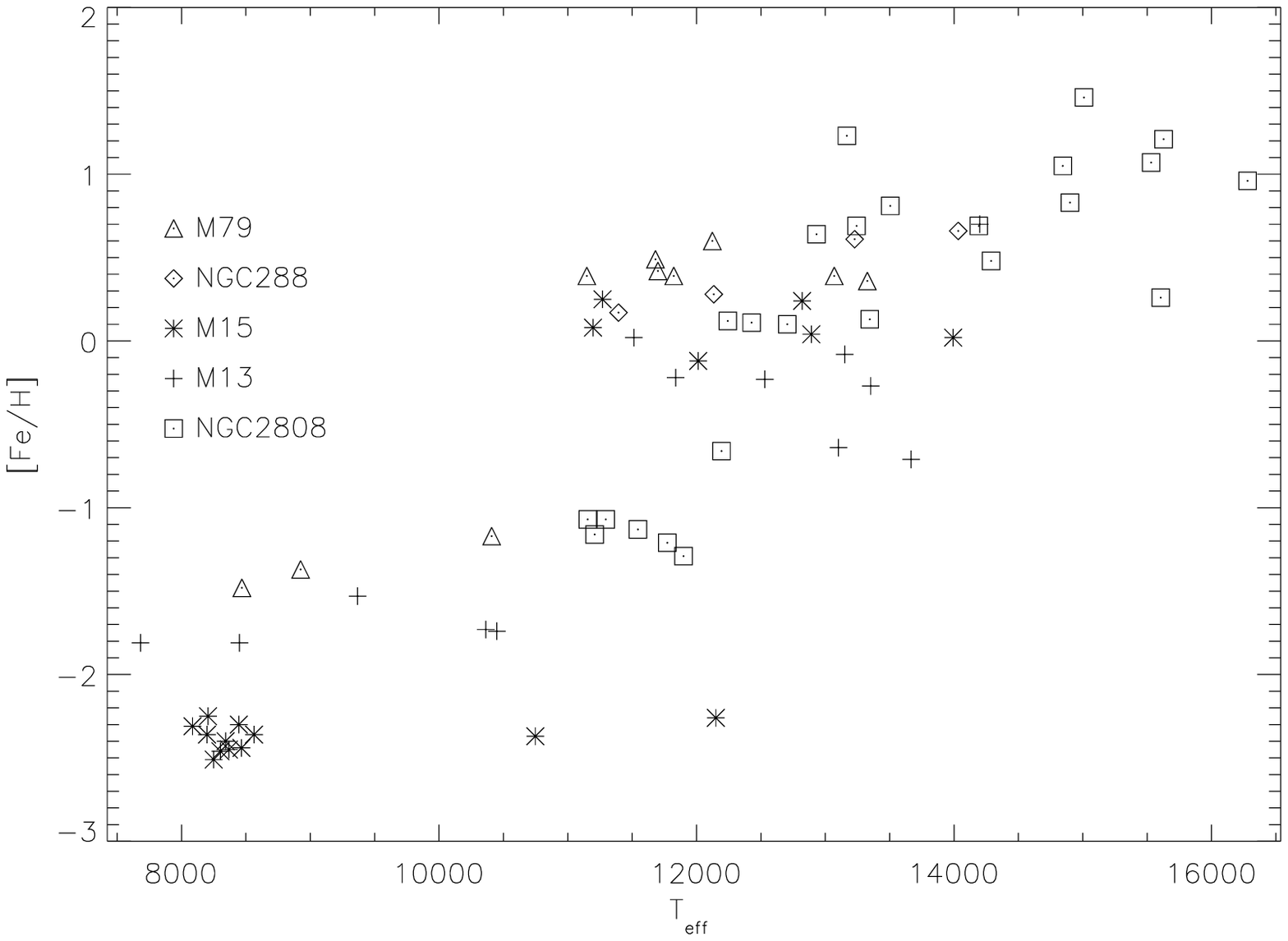}\\
\includegraphics[width=8cm]{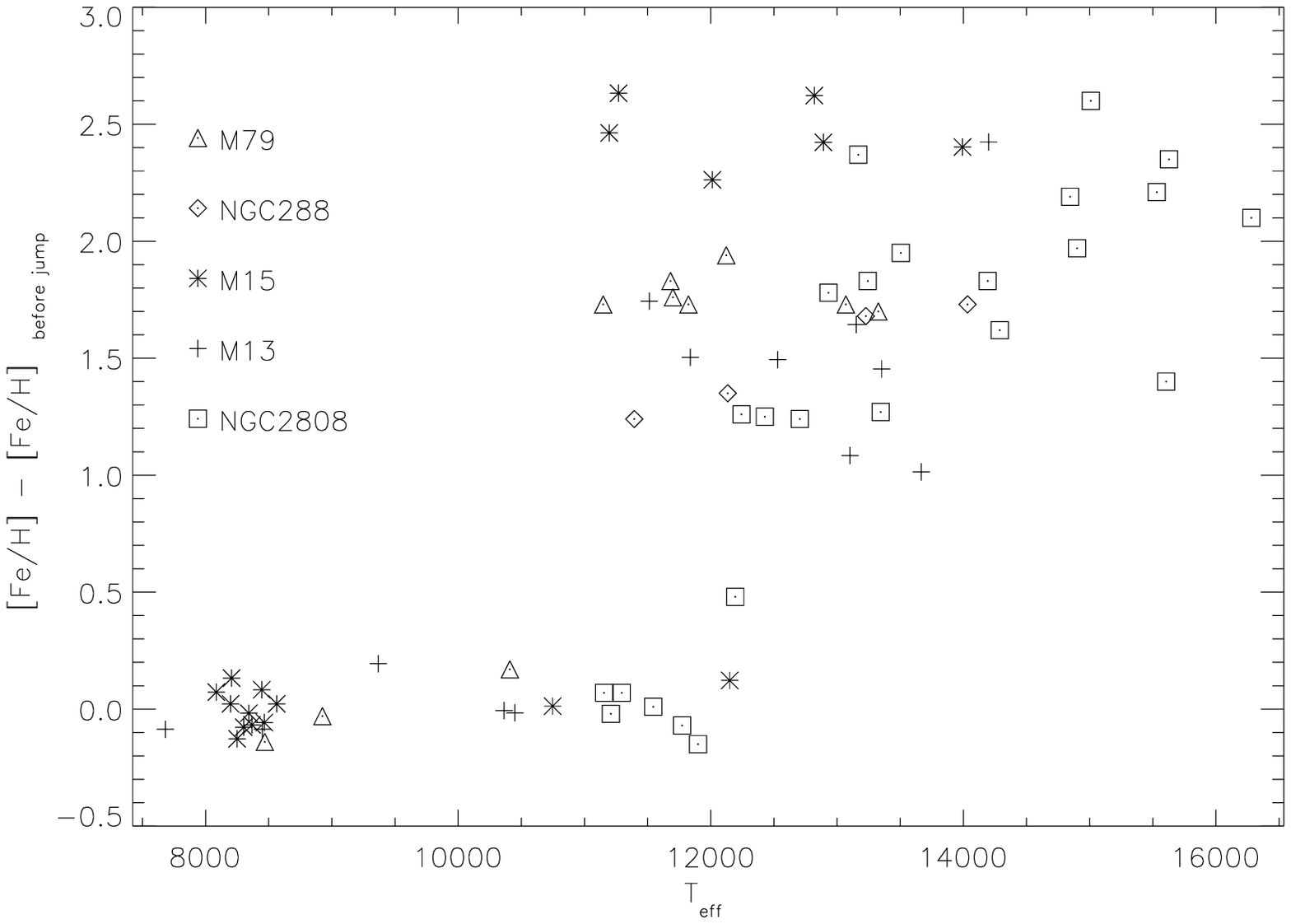}\\
\end{tabular}
\caption{
Comparison of the iron  abundances in blue  HB NGC~2808 stars with the
literature values   of   M~15, M~13,  NGC~288  (Behr,  2003)  and M~79
\cite{damian}.
 }
\label{fig5}
\end{figure}

\begin{figure}[h]
\centering
\begin{tabular}{c}
\includegraphics[width=8cm]{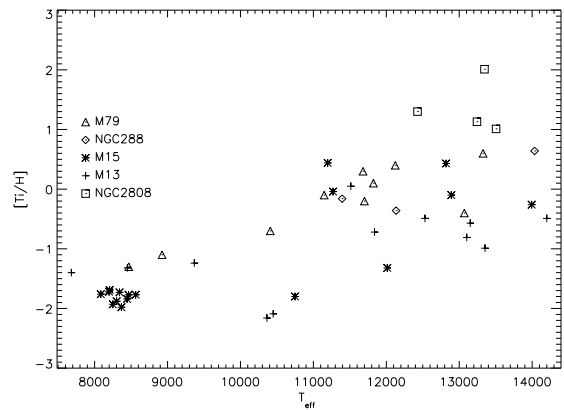}\\
\includegraphics[width=8cm]{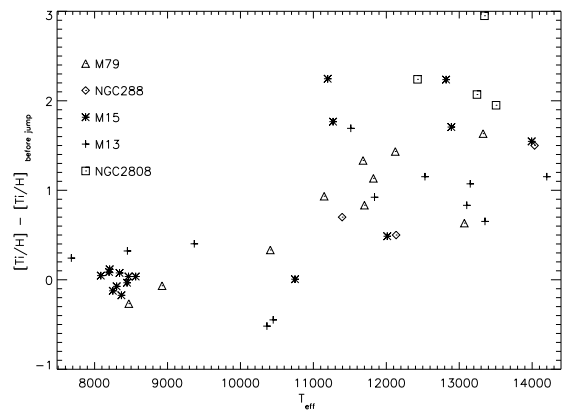}\\
\end{tabular}
\caption{
Same as Fig.~2 but for the titanium abundances }
\label{fig6}
\end{figure}

\begin{figure}[h]
\centering
\begin{tabular}{c}
\includegraphics[width=8cm]{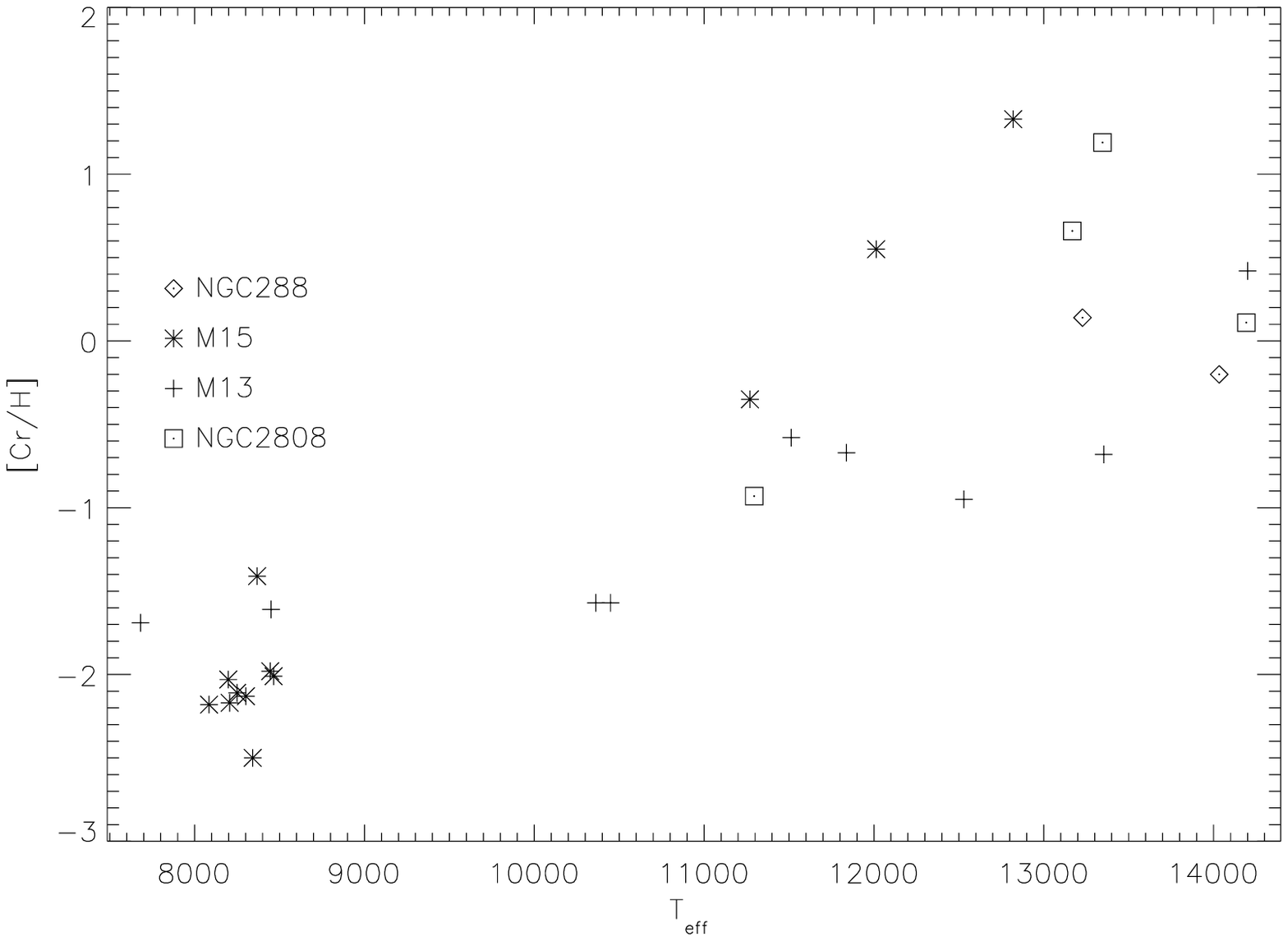}\\
\includegraphics[width=8cm]{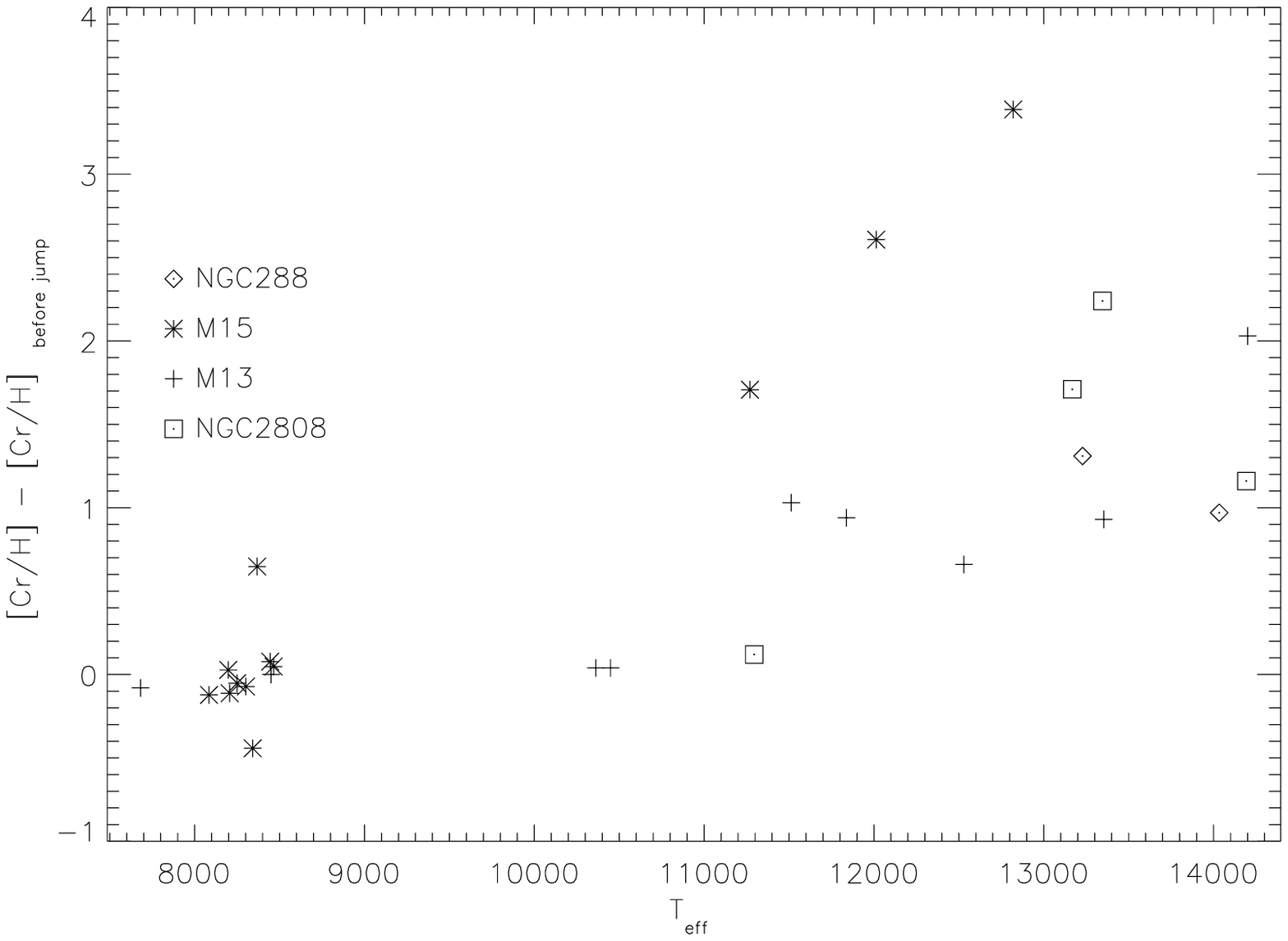}\\
\end{tabular}
\caption{
Comparison of the chromium abundances.}
\label{fig7}
\end{figure}


Due to the  limited S/N ratio and  the presence of cosmics, detections
of chromium and titanium lines with a shape  smooth enough to reliably
measure their equivalent width, could only be made for 4 stars.

\section{Comparison with other clusters}

The   measurement of radiative levitation  in  clusters with different
metallicities  is necessary  to provide the  appropriate  data for the
development  of  the  models.    NGC~2808  offers  the opportunity  of
observing this effect   in  a  cluster with   a   very  particular  HB
morphology and a younger  age  \citep{franciui}.

An important cross check of the ubiquitous nature of the phenomenon of
the levitation comes also from the comparison with NGC~288, a cluster
with a metal content very similar to that of NGC~2808.

\subsection{Metal enhancements and cluster metallicity}

Figure~\ref{fig5}, upper panel,  shows the iron abundances derived  in
this  paper as a function of  the effective temperature, together with
those of \citet{behr1} for  M~15, M~13 and  NGC~288, and those of M~79
from \citet{damian}.  Error bars are  not  shown in  the Figure for  a
purpose  of  clearness.    The same  data   are  also  shown in  Table
\ref{tabCFRspread}, where we   have written the   abundance range that
iron, chromium and  titanium spread in  five  clusters analysed.  The
clusters in the Table are  sorted according to their abundance nominal
value, from the most metal poor  in the top  to the most metal rich in
the bottom. It can be  noticed, looking at the  data about iron,  that
the   most  metal  poor  clusters  tend    to  have  larger  abundance
enhancements, suggesting the idea that  all the clusters tend to reach
the same value  after the   jump.  The  abundance  spread within  each
cluster  after the jump is very  high, much more  than for the coolest
stars, because of the trend with temperature, at  least in the case of
NGC~2808, and possibly because of larger  errors.  As a consequence we
do  not see a   narrower abundance distribution   of the stars  of all
clusters after the jump. This trend  is not confirmed for titanium and
chromium,  which, however, have  a poorer statistic. In any case, also
for iron, no firm conclusions can be stated at the moment.

Another apparent feature of   Figure~\ref{fig5} is the overlapping  of
non--enhanced and  enhanced region of  the  various clusters.  This is
likely due to error  in temperature determinations,  which, as we have
seen  in Section  3,  are  quite large   for  a given sample   studied
homogeneously.   They must  be    considerably larger  when  comparing
clusters analysed  by   different  authors, since, as     discussed in
Section~3,  not only differences in photometric  zero point are there,
but also the way  of   computing temperatures  out of  the  photometry
relies on  models that do   not match the  data  points.  It therefore
certainly introduces errors, which  could differ from study to  study.
We   cannot  exclude, however, the   possibility   that the  onset  of
radiative levitation takes place at slightly different temperatures in
different clusters.  Except   for  the hottest non--enhanced star   in
M~15, NGC~2808  has the  hottest onset  of levitation  in  any of  the
clusters in Figure~\ref{fig5}.  Once again, whether  this is real and,
in this    case,  what other properties   are   related to   the onset
temperature of the levitation, are questions that can only be answered
with sufficient  confidence when models that do  match the data points
become available.

\begin{table*}
\begin{center}
\begin{tabular}{c c c c c c }
\hline
\hline  
Cluster&\multicolumn{2}{c}{Before jump}&\multicolumn{3}{c}{After jump}\\
&[M/H] range&N$_{stars}$&[M/H] range&$\Delta$[M/H] range&N$_{stars}$\\
\hline
\multicolumn{6}{c}{IRON}\\
        M 15 &  -2.51/  -2.25 & 12 &  -0.12/   0.25 &  2.24/   2.61 &  6\\
        M 13 &  -1.81/  -1.53 &  5 &  -0.71/   0.70 &  1.03/   2.44 &  8\\
        M 79 &  -1.48/  -1.17 &  3 &   0.36/   0.60 &  1.70/   1.94 &  7\\
    NGC 2808 &  -1.29/  -1.07 &  6 &  -0.66/   1.46 &  0.50/   2.62 & 18\\
     NGC 288 &       -1.07    & -- &   0.17/   0.66 &  1.24/   1.73 &  4\\
{\bf Total}  &  -2.51/  -1.07 & 26 &  -0.71/   1.46 &  0.50/   2.62 & 43\\ 

\hline
\multicolumn{6}{c}{TITANIUM}\\  
         M 15 &  -1.98/  -1.69 & 11 &  -0.26/   0.44 &  1.53/   2.23 &  5\\
         M 13 &  -2.16/  -1.24 &  5 &  -0.99/   0.05 &  0.41/   1.45 &  7\\
         M 79 &  -1.30/  -0.70 &  3 &  -0.40/   0.60 &  0.63/   1.63 &  7\\
     NGC 2808 &       --       & -- &   1.01/   2.01 &  1.95/   2.95 &  4\\
      NGC 288 &       --       & -- &  -0.36/   0.64 &  0.50/   1.50 &  3\\
{\bf Total}   &  -2.16---0.70  & 19 &  -0.99/   2.01 &  0.41/   2.95 & 26\\ 
 
\hline
\multicolumn{6}{c}{CHROMIUM}\\

         M 15 &  -2.50/  -1.41 &  9 &  -0.35/   1.33 &  1.71/   3.39 &  3\\
         M 13 &  -1.69/  -1.57 &  4 &  -0.95/   0.42 &  0.70/   2.07 &  5\\
     NGC 2808 &       --       & -- &  -0.93/   1.19 &  0.12/   2.24 &  4\\
      NGC 288 &       --       & -- &  -0.20/   0.14 &  0.97/   1.31 &  2\\
{\bf Total}   &   -2.50---1.41 & 13 &  -0.95--  1.33 &  0.12/   3.39 & 14\\
\hline
\end{tabular}
\end{center}
\caption{
Comparison of the abundance ranges before and after the jump. The data
about M~79 are from Fabbian et al. (2005), those  about M~13, M~15 and
NGC~288 are from Behr  (2003a).  For the stars after  the jump we also
show   the range of  values spread  by the abundance   jumps, i.e. the
enhancement of the star abundance with respect to the nominal value of
the  cluster ($\Delta$[M/H]).  In  Behr's sample, NGC~288 has no stars
before the jump, so the metallicities in the  Table are those from RGB
stars, by Gratton (1987).  In the same way, since  we could not detect
titanium and  chromium  lines  in   stars before  the   jump, we  used
measurements in RGB stars by Carretta (2005).}
\label{tabCFRspread}
\end{table*}

\subsection{The case of NGC~2808}

If  we  now examine the  measurements  in NGC~2808,  it   is clear, as
pointed out in  Section 4, that a pattern  of  radiative levitation is
also present for blue HB   stars.  Nevertheless, the iron  enhancement
for  stars between $\sim$  11\,500~K and 13\,000~K, seems smaller than
what  is found in other clusters.   Even  NGC~288, with a very similar
metallicity to that  of NGC~2808,   but with  only 4  stars  analysed,
apparently   shows a higher   influence of radiative levitation, right
after the  jump, than NGC~2808  stars.  No measurements of titanium or
chromium are available for NGC~2808 in that range of temperature.

On  the  other hand,  the hotter stars   analysed in  NGC~2808 present
important signs of radiative levitation, arriving to iron enhancements
slightly smaller than   those  of M~15 at  lower    temperatures.  The
enhancements of  titanium and chromium  for stars with T$_{\rm eff} >$
13\,000~K are  also very remarkable  and  compatible or  even slightly
higher than those observed in other clusters.

In brief, NGC~2808 seems to have a  progressive metal enhancement with
temperature after T$_{\rm eff}$ $\sim$   12\,000~K, spanning a  larger
range  of abundances than  other clusters, for  which  there is a more
abrupt discontinuity at the level of the  jump.  Is this behaviour due
to  the higher   metallicity of  NGC~2808, to   the complexity of  its
horizontal branch,   or,   more simply, to  the   better  coverage (in
temperature) of the HB in our analysis?  In this sense, the comparison
with   NGC~288,  having a similar  metal  content  but a  different HB
morphology and age, can be   very helpful.  Unfortunately, up to  date
only four stars have been analysed in NGC~288  and they are not enough
to perform  an adequate comparison  of the abundance anomalies in both
clusters.  In  the future,  it  would be very  interesting  to put  in
relation the couple  of clusters NGC~288  - NGC~2808, at a metallicity
of [Fe/H] $\sim -1.1$,  with  the couple  formed  by M~79 and M~13  at
[Fe/H] $\sim -1.4$.

\section{Conclusions}

The  objective of this   investigation was to  shed more  light on the
nature  of radiative   levitation   and diffusion of  elements   among
horizontal  branch  stars   and   its connection with   the   physical
properties of the  affected  stellar population.  Globular  cluster HB
stars come from a coeval population with a very small, if any, initial
abundance spread. As a consequence they offer  a unique opportunity of
tackling this issue.  Studies, including  the present, made so far  on
the metallicity of HB stars, lack of appropriate models describing the
atmospheres of  the enhanced  stars.   Abundance measurements  at this
stage, however, point to  important conclusions which are unlikely  to
be a   mere  artefact of  the   measurement method: the   slope in the
temperature vs.  abundance  diagram is  significantly higher than  the
errors involved  and the metal content  of the cluster  has probably a
primary role in determining the amplitude of the jump (more metal poor
clusters  show more enhancement among stars  after  the jump).  To our
knowledge,   model atmosphere  able  to   account   for the  radiative
levitation are  not yet available, even  though a  theoretical work on
the phenomenon  has been  made (see Section  \ref{intro}).  If, on one
hand,  the analysis here presented  will be considerably improved when
more adequate model atmospheres can be  adopted, on the other hand the
achievement proper  models  require observational  clues regarding the
amount of metal enhancement and its  dependence on physical parameters
and clusters' properties.   This paper is also  meant to contribute to
the collecting of such observational clues.

\begin{acknowledgements}
We owe special thanks to Santino Cassisi, for  he provided us with the
HB--stars evolutionary tracks   and for  very useful  discussion   via
e--mail.  In order to retrieve the line list necessary to the chemical
analysis,  we have  used  the Vienna Atomic  Line  Data--base.  We are
greatly indebted  to  Raffaele Gratton  and Damian  Fabbian for having
made available to us their code for the abundance measurements and for
exhaustive explanations.  We also  thank Patrick Francois for allowing
us to use his code for  the automatic detection and equivalent width
measurement of the lines,  even though,  due to  the faintness  of the
stars  studied,  a more  interactive  approach  had  to be  eventually
adopted.   Marco Montalto spent time and   patience helping one of the
authors with   a bash  script.  The  paper  benefits  from the  useful
comments of the  anonymous  referee.  G.c.\ P.,   G.p.\ P., \&  Y.\ M.
acknowledge the support of italian MIUR under the programme PRIN 2003.
A.  R.--B.  acknowledges the support of the European Space Agency.

\end{acknowledgements}

\bibliographystyle{aa}

\end{document}